\documentclass[final,5p,times]{elsarticle} 

\usepackage{graphicx}
\usepackage{amssymb}
\journal{Nuclear Instruments and Methods in Physics Research, Section A}

\usepackage{url}
\usepackage{xspace}
\newcommand {\sci}[2]    {\mbox{\ensuremath{ #1 \! \cdot \! 10^{#2} }}}
\newcommand {\scix}[1]   {\mbox{\ensuremath{ 10^{#1} }}}
\newcommand {\defuni}[1] {\ifmmode \mathrm{#1} \else $\mathrm{#1}$ \fi}
\newcommand {\um}[1]     {\defuni{\; #1}}
\newcommand {\DD}[1]     {\ensuremath{\mathinner{\Delta#1}}}
\newcommand {\degr}      {\ensuremath{^{\circ}}}
\newcommand {\EUSO}      {$\mathcal{EUSO}$\xspace}
\begin{document}

\begin{frontmatter}



\title{Ultra-High Energy Cosmic Particles studies from space: \\ super-\EUSO, a possible next-generation experiment.}


\author{Alessandro Petrolini}

\address{
Dipartimento di Fisica dell'Universit\`a di Genova e INFN sezione di Genova,\\
Via Dodecaneso 33, I-16146, Genova, Italia.
}

\ead{Alessandro.Petrolini@ge.infn.it}

\begin{abstract}

After the Pierre Auger Observatory, it is likely that space-based experiments
might be required for next-generation studies of Ultra-High Energy Cosmic
Particles.  An overview of this challenging task is presented, emphasizing the
main design issues, the criticalities and the intermediate steps
required to make this challenging task a reality.

\end{abstract}


\end{frontmatter}

%%% \linenumbers

%-------------------------------------------------------------------------------
\section{ Introduction }
%-------------------------------------------------------------------------------

The interpretation of the phenomenology of Ultra-High Energy
Cosmic Particles (UHECP) is one of the most challenging
topics of modern astro-particle physics. 

UHECP
reach the Earth with a very low directional intensity of
a few particles $\um{millennium}^{-1}\um{km}^{-2} \um{sr}^{-1}$
for particles with an energy $E \gtrsim 10^{20} \um{eV}$)~\cite{bi:PAOSpectrum} and
therefore a large, complex and sophisticated
experimental apparatus is required to observe them.

The science case of UHECP will not be
presented in this paper as it has been extensively discussed
elsewhere (for instance
see~\cite{bi:WhitePaper,bi:APPECRoadMap,bi:SEUSO,bi:Focus} an references
therein). 
This paper will only discuss the experimental
apparatus required for space-based experiments for the observation of UHECP 
in the post Pierre Auger Observatory (PAO) era: perspectives,
how-to-do and a road-map will be identified.

I assume the point of view that, after the PAO south site, which is the present
state-of-the-art experiment in the field, the hopefully near future is the full
PAO observatory, with the north site added to the already operating Auger-south site.
Moreover I assume that the, hopefully not too far, future might be a
space-based experiment.

%-------------------------------------------------------------------------------
\section{The observational technique}
%-------------------------------------------------------------------------------

\begin{figure}[htb]
 \begin{center}
   \includegraphics[width=0.50\textwidth]{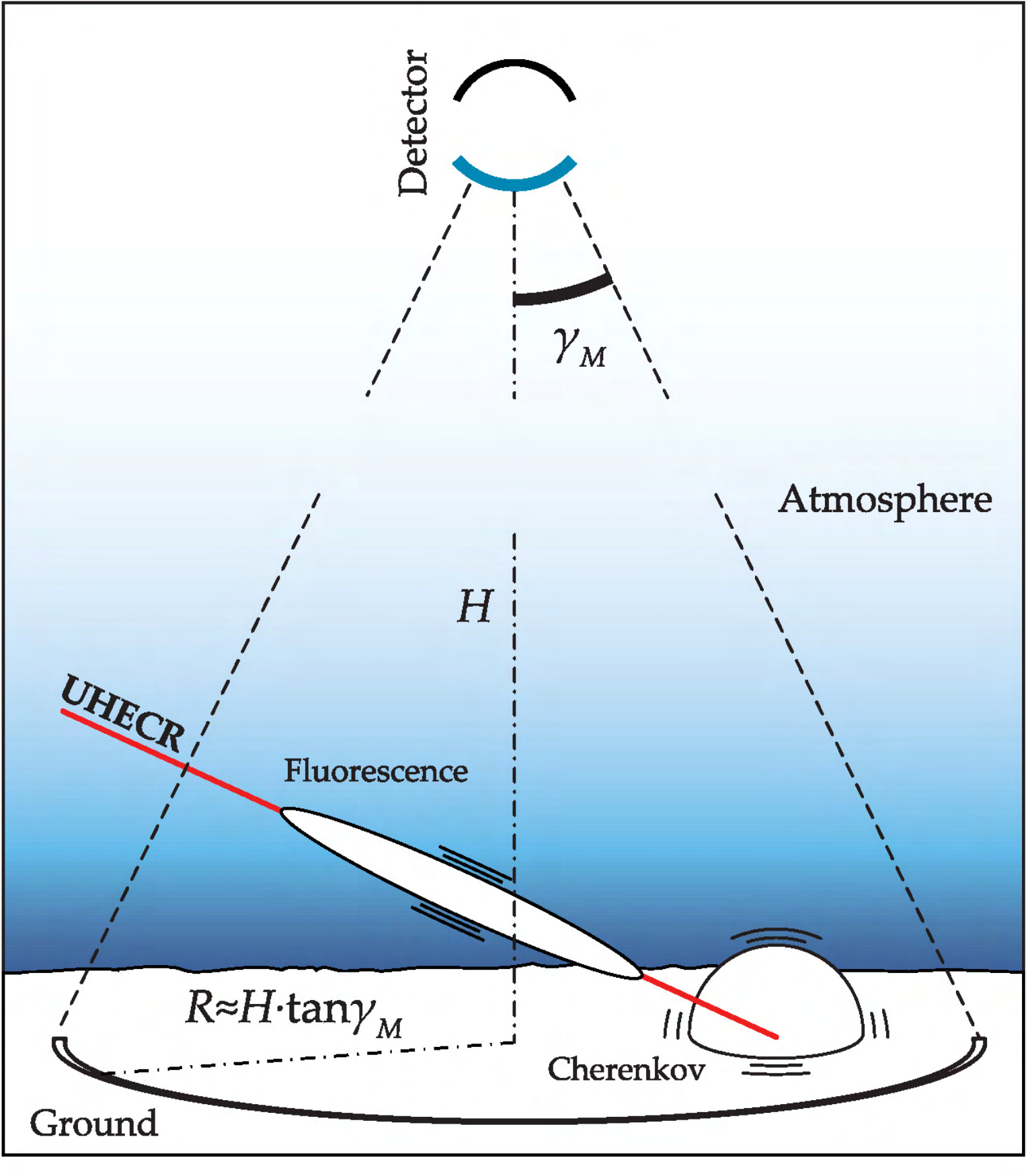}
   \label{fi:EUSOApproach}
 \end{center}
\caption{ The AirWatch observational approach.}
\label{fi:approach}
\end{figure}

\EUSO and super-\EUSO are implementations of the \textit{AirWatch concept}, originally
proposed by John Linsley more than twenty years
ago~\cite{bi:JL,bi:EUSORedBook}: to observe from space the Extensive Air Shower (EAS) produced
by the interaction of a primary cosmic particle with the atmoshpere, as
shown in Figure~\ref{fi:approach}.

An EAS can be detected by observing the air scintillation light,
produced during the EAS development.
The additional observation of the diffusely reflected
Cherenkov light (reflected either by land, sea or clouds) provides
additional information. 
It is then possible to estimate the
energy and arrival direction of the primary UHECP and to gather
information about its nature. 
The atmosphere is used as a calorimeter.
Any given EAS will be seen as a point moving with a
direction and angular velocity depending on the EAS direction. 

The air scintillation light is mainly emitted in the wavelength range 
\mbox{$330 \um{nm} \div 400 \um{nm}$}, concentrating around three bands at
$\lambda_1\approx 340\um{nm}$,
$\lambda_2\approx 360\um{nm}$ and  
$\lambda_3\approx 390\um{nm}$. 
It is isotropic and proportional, at any
point along the EAS development, to the number of charged particles in the EAS, largely
dominated by electrons and positrons.
The total amount of light produced is proportional to the primary
particle energy and the shape of the EAS profile (in particular the
atmospheric depth of the EAS maximum) contains information about the
primary particle identity.
In this wavelength range the atmosphere is relatively transparent, down to $\lambda \approx 330 \um{nm}$
where the ozone absorption becomes strong.

The possible observation of the Cherenkov light diffusely reflected by
the Earth (by land, sea or clouds) will help the determination of the EAS parameters. 
While the amount of observed Cherenkov photons depends
on the reflectance and geometry of the impact surface, the
directionality of the Cherenkov beam provides a precise extrapolation
of the EAS direction to the first reflecting surface.

A selective and efficient trigger is required to distinguish the EAS
from the background.  
Many different kind of backgrounds are expected including: man-made
lights, auroras, natural photo-chemical effects (in atmosphere, sea
and land), low-energy cosmic rays, reflected moon-light and
star-light as well as the most important expected contribution: the random
night-glow background.
The random night-glow is currently estimated in the range
$B\approx\sci{ (3\div 15) }{11} \um{photons \cdot m^{-2} \cdot s^{-1} \cdot sr^{-1}}$,
in the wavelength range
\mbox{$330 \um{nm}\le\lambda\le400\um{nm}$} at $ H \approx 400 \um{km} $
height, depending on various factors.

The peculiar characteristics of the EAS, including the kinematical
ones, allow one to distinguish them from the various backgrounds,
because those have a typically different space-time development.
The non-random backgrounds, for instance, typically have a
time-scale of the order of \um{ms}, much longer than an EAS.

The challenging goal can be accomplished from space by
observing the Earth atmosphere at night-time, looking down to nadir.

Key points of the observational technique are the following.

\begin{enumerate}

\item
A large instantaneous geometrical aperture can be obtained, thanks to the large
distance, depending on the Field of View (FoV) of the apparatus.
A large mass of atmosphere (the calorimeter medium) can therefore be
observed.
\item
Detection of EAS produced by weakly interacting primary particles,
starting to shower deeply in the atmosphere, is possible, by the 
direct observation of the EAS development and starting point.
\item
All sky coverage is possible with one single apparatus.
\item
The approach is complementary to the ground-based one. In fact 
space experiments are best suited for the observation of higher energy
UHECP with respect to ground-based experiments. However an overlap of
the observed energy spectrum with the one known from ground-based experiments
is required for a safe comparison. 
The systematic effects are largely different in the two approaches.

\end{enumerate}

%-------------------------------------------------------------------------------
\section{ Beyond the Pierre Auger Observatory from space ? }
%-------------------------------------------------------------------------------

Exceeding the PAO performance requires a huge experiment, on a long time-scale,
requiring a large amount of preliminary R\&D and ancillary studies.

Morevoer the challenging goal of a big experiment from space requires intermediate steps
some of which are on the way (JEM-\EUSO,~\cite{bi:JEMEUSO}).
Other intermediate steps include some path-finder and/or technological model, as
it will be discussed later.

What we are learning from PAO will help to tune the future scientific objectives.
A tuning of the scientific objectives is mandatory because 
the Experiment-For-Everything is, most likely, impossible: 
a choice of scientific objectives will be needed, 
to drive trade-offs and choices on the experiment design and performance.

Such a challenging enterprise would require the involvement
of a large part of the UHECP physicists community, in a coordinated effort.

Science must be, obviously, the driving force.
However, after a careful assessment of the predictable status of UHECP science 
in some ten/fifteen years from now, based on the exploitation of the full PAO experiment,
one needs to consider a realistic implementation of possible experimental apparata,
to avoid dreams which will never become a reality, 
at least on the time-scale of a human life.

Most new big projects have a long time duration
and require a long time for conception, design, commissioning (plus getting funds...).
Therefore one must start right now to think about concrete proposals
for realistic implementations of post PAO (south+north) experiments.
It is not too early, as some ten years are certainly necessary to design and
build such a challenging apparatus.

%-------------------------------------------------------------------------------
\section{The apparatus required for space-based UHECP observation}
%-------------------------------------------------------------------------------

The AirWatch observational technique was carefully described in~\cite{bi:EUSORedBook}.
The required apparatus is an Earth-watching 
large aperture, large Field-Of-View (FoV), fast and highly pixelized 
digital camera for detecting near-UV single photons
superimposed on a huge background and
capable of (at least!) five years operation in space~\cite{bi:SpacePaper}.

It is made of:

\begin{enumerate}

\item
a main optics, collecting photons and focusing the EAS image
onto its Focal Surface (FS);

\item
the photo-detector (PD) on the FS, for registering the EAS image, which is made of: 
photo-sensors, front-end electronics, back-end electronics, triggering and data analysis systems;

\item
possible ancillary instrumentation such as: 
\begin{enumerate}
\item
a LIDAR for atmospheric monitoring;
\item
an IR camera;
\item
a suitable radio-detection system .
\end{enumerate}

\item
system instrumentation.

\end{enumerate}

%-------------------------------------------------------------------------------
\section{Some order of magnitude estimates}
%-------------------------------------------------------------------------------

Some order of magnitude estimates are derived in this paragraph,
in order to provide figures which are necessary for understanding the concept design.

Assume an apparatus looking down the Earth at night from $ H \simeq 400 \um{km}$ height, 
with an entrance pupil diameter $ D \simeq 5 \um{m} $, 
and half-angle FoV $\gamma= 20\degr$ and a 
total apparatus detection efficiency $\epsilon \approx (0.1 \div 0.2)$.

%-------------------------------------------------------------------------------
\subsection{The EAS signal}
%-------------------------------------------------------------------------------

Consider an hadron-induced EAS 
with energy
$ E \simeq 10^{19}\um{eV} $, 
zenith angle
$\theta_z \simeq \pi/4$, 
and azimuth angle with respect to the radial direction on the FoV 
$\phi=\pm \pi/2$.

Under such hypoteses the time-integrated signal reaching the apparatus
(irradiance) is 
$ \mathcal{I} \approx 50 \um{ photons \cdot m^{-2} } $, which means that a large entrance pupil is required.
This irradiance implies a total number of detected photons from the EAS given by
$\mathcal{I} A \epsilon \approx 150 $, which translates into an energy resolution
$\DD{E}/E \approx 0.1$ (due to the statistical error only).

The apparent time duration of the visible EAS track is $T \approx 80 \um{\mu s}$
and its typical angular span is $ \DD{\xi} \approx 1.5 \degr$.
These facts imply that a time resolution at the $\um{\mu s}$ level and a fine
granularity of the photo-detector (better than $ \approx 0.1\degr
\times 0.1\degr $) are needed in order to get an angular resolution on the primary EAS particle
direction of $\DD{\theta} \approx 1\degr$.
This fact and the large FoV require a large number of channels (of the order of
$ \scix{5} \div \scix{6} $).

%-------------------------------------------------------------------------------
\subsection{The random night-glow background}
%-------------------------------------------------------------------------------

In the real world background makes the previous estimations too
optimistic~\cite{bi:SpacePaper} as the large background must be accounted for.

The reference random night-glow background in the $330\um{nm} \div 400 \um{nm}$ wavelength range is
$B\approx\sci{5}{11}\um{photons \cdot m^{-2} \cdot s^{-1} \cdot sr^{-1}}$ but
current estimates show that it might be be up to a factor two larger~\cite{bi:EUSORedBook}. Morevoer
its detailed space-time structure, at the level of the characteristic space-time
scales of the EAS development, is not known.

The resulting total random night-glow background rate intercepted (on the whole entrance pupil and
full FoV) is then $ \approx 4 \um{THz}$.

The resulting total random night-glow background rate detected on the PD per pixel 
(pixel size: $ 0.1\degr \times 0.1\degr $ for a total of $N = \sci{1.2}{5}
\um{pixels}$) is then $ 3 \um{MHz/pixel} $.

These figures result in one order of magnitude more random night-glow background than signal photons
superimposed on the typical EAS (over all its space-time development) as well as 
roughly the same number of signal and random night-glow background photons near the EAS
maximum. 
Therefore one needs to find a way to cope with the huge background, taking into
account that many other sources of non-random background come into play. 

%-------------------------------------------------------------------------------
\subsection{Extraction of the EAS signal from the random night-glow background}
%-------------------------------------------------------------------------------

The previous figures show that it is a real challenge to extract the EAS signal
from the random night-glow background.
This is particularly true if one aims to go down safely in the UHECP energy to 
$E \approx \scix{19} \um{eV}$, 
as good physics would demand. In fact the signal and random
night-glow background scale proportionally with the entrance pupil area with a
signal decreasing when looking at EAS of progressively decreasing energy 
over a constant random night-glow background.

In order to extract the EAS from the background 
a precise knowledge of the properties of the background is required,
at the space-time level of the EAS kinematics.
A continuos monitoring (and subtraction) of the average background
on a pixel-by-pixel (or so) basis is thus unavoidable to decrease the energy
threshold down to $E \approx \scix{19} \um{eV}$.

The acceptable background level also depends on the energy of the EAS.
In order to allow for background dependent observations 
a precise knowledge of the apparatus sensitivity as a function of the background is required,
which requires a precise apparatus calibration.

In order to face the background a clever and powerful triggering and data-handling scheme 
need to be invented.

Many other types of backgrounds, in addition to the random night-glow one, exist.
Most of them appear not to be dangerous as their kinematics very different from
the one of an EAS,
but it is most likely mandatory a dedicated background measurement to improve
our knowledge.

%-------------------------------------------------------------------------------
\section{\EUSO}
%-------------------------------------------------------------------------------

\EUSO~\cite{bi:EUSORedBook} was proposed to the European Space Agency (ESA) as a free-flyer with:
a main optics diameter of $ \simeq 3.5 \um{m}$;
orbit height and inclination as free parameters to be tuned;
mass, volume, power, telemetry and other budgets largely unconstrained.

ESA recommended to consider the accommodation on the ESA Columbus module on the
International Space Station (ISS) so that many constraints had to be taken into account, including: 
a fixed orbit and 
severe limits on mass, volume, power, telemetry and other budgets.
In fact the accommodation had to be made compatible with the ISS/Columbus resources including: 
mass ($ 1.5$ ton), 
volume ($2.5 \times 2.5 \times 4.5 \um{m^3}$), 
power ($1\um{kW}$) and telemetry ($180 \um{Mbit/orbit}$).
The volume, in particular, was limited by the envisaged accommodation on the
Space Shuttle, not by its
capabilities.

Many constraints, mainly due to the ISS, limited the final \EUSO performance
which was finally put into hibernation by ESA.

%-------------------------------------------------------------------------------
\subsection{ The heritage of \EUSO projected into the future }
%-------------------------------------------------------------------------------

\EUSO underwent a detailed phase-A study.

We have learnt a lot about space-based observation of UHECP
from the \EUSO phase-A studies.
The \EUSO heritage is exceedingly precious and it is wise
to exploit what we learnt from \EUSO
to develop a second generation experiment.

We can consider \EUSO as our prototype exercise in the conception and design of
which we have possibly done mistakes!

One of the most important lessons is that one needs a safe design margin on the
expected performance already at a design stage.

%-------------------------------------------------------------------------------
\section{ From \EUSO to super-\EUSO }
%-------------------------------------------------------------------------------

Aiming at (or dreaming of) an experiment with much better performance than \EUSO
the following consequences (all relative to the \EUSO
design~\cite{bi:EUSORedBook}) must be faced.

The \EUSO efficiency plateau was reached for energies larger than  
$E \approx \sci{( 1 \div 2 )}{20} \um{eV}$:
one needs to gain a factor $( 10 \div 20 )$ - at least - in the energy threshold
in order to cross-calibrate with ground-based exepriments, to safely observe the
GZK region and increase the number of expecteed neutrino events.

In order to observe fainter events one needs to collect more photons.
More collected photons improve the energy resolution and angular resolution on
the UHECP as well.
However increasing the number of collected photons also increases the background rate: 
it is not enough to increase the number of photons to improve the performance.
In fact at some low-enough energy the EAS will fade away in the background.

The FoV of the optics might be reduced to get better optical efficiency: 
assume $ \gamma_M = (20\degr\div 25\degr)$ (half-angle).

After that one has to recover the instantaneous geometrical aperture by higher orbits.

The number of channels is a challenge anyway: if one assumes a maximum of one
million channels (it is already very challenging) one finds a angular
granularity of $0.06\degr$ (one mrad).

As one needs to account for a higher orbit, this requires smaller pixels to
obtain the same granularity at the Earth.

%-------------------------------------------------------------------------------
\subsection{Optimisation of the orbital parameters}
%-------------------------------------------------------------------------------

A free-flyer allows many more degrees of freedom in the choice of the orbit than the ISS.
One may also think to change the orbit during the mission.
Changing the orbit height actually shifts up/down the observational energy range.
One can therefore consider tuning of the orbit: either elliptic orbit or orbit altitude change.

A reasonalbe baseline would be an elliptical orbit with perigee at $400 \um{km}$
and apogee at $1100\um{km}$, taking into account the space environment and the
experiment requiremetns.

A tilting of the apparatus in order to increase the instantaneous geoemtrical
aperture is not required. The effect of titlting was carefully described in~\cite{bi:SpacePaper}.

It should be stressed that a free-flyer might allow for repetitive passes above
specified locations at the Earth (for observing calibration sources, and,
possibly for cross-calibration with the PAO sites).

%-------------------------------------------------------------------------------
\subsection{ Technical Developments: from \EUSO to super-\EUSO }
%-------------------------------------------------------------------------------

A number of technological developments are indeed mandatory in order to realize
the super-\EUSO apparatus, including:
a deployable catadioptic optics (such that a much larger optics is possible);
use of better photo-sensors, such as GAPD, (having much better characteristics
than photomultiplier-based sensors, including a better expected total detection efficiency);
develpoment of low-power instrumentation (sensors, front-end electronics,...);
developement of better optical filters allowing separation of the nitrogen
scintillation signal from the almost uniform random night-glow background;
lightweight materials with low chromatic aberrations for the transapretn
components of the optics.

All these technological developments are interesting for many other applications so that funding for R\&D
as well as spin-off is possible.
Moreovoe these R\&D are inside a technological road-map, a fact whicih will help
to push the scientific objectives.

%-------------------------------------------------------------------------------
\subsection{ Expected performance~\cite{bi:SpacePaper} }
%-------------------------------------------------------------------------------

Provided suitable technologies are successfully developed the following
performance paramters can be reached.

Area and instantaneous geometrical aperture (depends on orbit and FoV):
$ \approx \sci{0.8}{6} \um{km^2} $ observed area (at aphelion);
$ \approx \sci{2}{6} \um{km^2 \cdot sr} $ instantaneous geometrical aperture (at aphelion).
The duty cycle is not included as mesurements are required to properly estimate
it.

Threshold energy depends on the optics entrance pupil and total photo-detector
efficiency (PDE).
However the optics aperture is the only sizeable parameter, to within the external constraints;
the PDE can realistically improve by a factor two or so (it is the only factor
much smaller than one);
a lot of other factors affect the overall efficiency, but all of them are
already close to one.

As a guess-estimate: with $ D \approx 6 \um{m} $ and PDE doubled with respect to \EUSO 
one can think to reach $ E \approx \scix{19} \um{eV}$ 
with $ \DD{E}/E \approx 0.1 $ (statistical error only).
But background subtraction is not trivial and must be worked out.

Moreover one can expect $ \DD{\theta} \approx 1\degr $ on the reconstructed primary particle direction,
limited by the EAS visible track-length and by the affordable number of channels.

On the basis of the previus reasoning the main super-\EUSO baselines parameters
and design goals are summarized in table~\ref{ta:SEUSO}.

\begin{center}
\begin{table*}[htp]
\begin{center}
\begin{tabular}{ll}
\hline
\multicolumn{2}{c}{\textbf{Main physical Parameters}}
\\
\hline
Operating Wavelength Range (WR)						
& 
	$\{330 \um{nm} \lesssim \lambda \lesssim 400 \um{nm}\}$
	\\
Background (in WR) at $\approx 750 \um{km}$ height.	
&
	$ \sci{(3 \div 15)}{11} \um{photons \cdot m^{-2} \cdot s^{-1} \cdot sr^{-1}} $	
	\\
Average atmospheric transmission (in WR)				
&   
	$ K_{atm} \gtrsim 0.4$ 
	\\
\hline
\multicolumn{2}{c}{\textbf{Orbital Parameters}}
\\
\hline
Orbit perigee								
& 
	$r_P \simeq 800 $ km
	\\
Orbit apogee								
& 
	$r_A \simeq 1100 $ km
	\\	
Orbit inclination							
& 
	$ i \approx ( 50\degr \div 60\degr ) $ 
	\\								
Orbital period								
&
       	$ T_0 \simeq 100 $ min						
	\\									
Velocity of the ground track                                            
&	
	$v_{GT} \simeq 7.5 $ km/s
	\\									 
Pointing and pointing accuracy					
&
	Nadir to within $\Delta \xi \simeq 3\degr$			
	\\								
\hline
\multicolumn{2}{c}{\textbf{Satellite Parameters}}
\\
\hline
Satellite envelope shape
&
	Frustum of a cone
	\\
Diameters
&
	$D_{MAX} \simeq 11 $ m and $D_{MIN} \simeq 7 $ m.
	\\
Length
&
	$ L \simeq 10 $ m
	\\
Operational Lifetime
&
	$ (5 \div 10) $ years
	\\

\hline
\multicolumn{2}{c}{\textbf{Main apparatus parameters and requirements}}
\\
\hline
Type
&
	Deployable catadioptric system
	\\
Main mirror
&
	$D_M \simeq 11$ m
	\\
Entrance pupil and corrector plate
&
	$D_{EP} \simeq 7$ m
	\\
Angular granularity
&
	$ \Delta \ell \approx 0.7 $ km at the Earth
	\\
Optics throughput
&
	$\epsilon_{O} \gtrsim 0.7 $
	\\
$ f/\# $
&
	$ \approx 0.6 $
	\\
Optics spot size diameter on the FS     
& 
	$ 3 \um{mm} \div 5 \um{mm} $   
	\\
Apparatus Field of View (FoV), half-angle:
&
	$\gamma_M = 20\degr \div 25\degr$
	\\
Total length of the optics
&
$ \approx 9 $ m
\\
Focal Surface size (diameter)
&
	$ \approx 4 $ m
	\\
PDE
&
	$ \epsilon_{PDE} \gtrsim 0.25$
	\\
Number of detector channels
&
	$ \approx 1.2 $ million
	\\
Size of the pixels on the PD
&
	$ \approx 4 $ mm
	\\
Photo-Sensor
&
	GAPD
	\\
Power consumption
&
	less than $ 2 $ mW per channel
	\\
\hline
\multicolumn{2}{c}{\textbf{Main Performance Parameters and requirements}}
\\
\hline
Low Energy Threshold
&
	$ E_{th} \approx \scix{19} \um{eV}$.
	\\
Instantaneous geometrical aperture
&
	$ A_G \approx \sci{2.0}{6} \um{km^2 sr}$ 
	\\
Statistical error on the energy measurement
&
	$ \Delta E / E \approx 0.1 \; @ \; E \approx \scix{19} \um{eV}$
	\\
Angular resolution on the primary direction
&
	$ \Delta \chi \approx ( 1\degr \div 5\degr)$
	\\
Observation Duty cycle
&
	$ \eta \approx ( 0.1 \div 0.2 )$
	\\
\hline
\multicolumn{2}{c}{\textbf{Main budgets} (at the present level of knowledge)}
\\
\hline
Mass
&
	5 ton
	\\
Power
&
	5 kW
	\\
Telemetry
&
	20 Gbit/orbit
	\\
\hline
\end{tabular}
\caption{The main super-\EUSO baselines parameters and design goals.}
\label{ta:SEUSO}
\end{center}
\end{table*}
\end{center}

%-------------------------------------------------------------------------------
\subsection{ Main intrisic critical issues of space-based observation }
%-------------------------------------------------------------------------------

The large distance implies that:
the signal is faint, 
angular resolution is not excellent and
the transverse extent of the EAS is not observable.

Non-random plus random (night-glow) backgrounds implies that the total
background is larger than from ground.
This huge background calls for an online subtraction for triggering.

Orbit optimisation should account for: obsearvational energy range, man-made
background, atmospheric phenomena and day-night effects as well as the very
large drag coefficient of such an apparatus requirign a long lifetime.

Stray-light control with such a large FoV and sensitive apparatus is a critical
desing issue.

Atmospheric monitoring is a critical and important item as
the observed FoV is continuously changing:
a continuous monitoring is needed
and parameter recording is required.

%-------------------------------------------------------------------------------
\subsection{ Main technological critical issues of space-based observation }
%-------------------------------------------------------------------------------

The main technological critical issues of space-based observation include:
the large deployable optics; optimal stray-light control of the large FoV and
highly sensitive apparatus;
the architecture, design and engineering of the
photo-detector with one million channels (front-end and back-end electronics
design and power, trigger design);
photo-detector protection from intense light (via attitude control and/or a shutter);
data-handling and detector calibration of $\approx$ one million channels on-orbit;
power consumption; mass; the need of a suitable (huge) launcher vehicle;
ageing of all the components in the harsh space environment;
protection from orbital debris of delicate instrumental parts;
engineering issues such as thermal control of the large volumes and surfaces and
critical mechanical parts; the need for a high-capacity battery system.

%-------------------------------------------------------------------------------
\section{ A large experiment for the ESA Cosmic Vision program (2015-2025)}
%-------------------------------------------------------------------------------

One possible opportunity to dream for a large future mission is 
the ESA Cosmic Vision program (2015-2025).
The study of UHECP from space entered the ESA road-map (so many other themes too...).

A new call for missions due for implementation in the period 2015-2025 is
expected to go out in 2010.
The previous selection (2007) judged the super-\EUSO proposal a scientifically valuable one
but technological readiness was judged very low.

All critical points detected by ESA were well known to the community.
They point out that a lot of work is required in order to prepare a new proposal.

%-------------------------------------------------------------------------------
\section{ The road-map and the intermediate steps }
%-------------------------------------------------------------------------------

The demanding requirements impact on resources.
This calls for a careful experiment optimisation in order to collect as much as
possible information at a preliminary stage
via a well defined road-map with intermediate steps.

In fact the challenging goal of a big space experiment does require intermediate steps,
some of which are on the way (JEM-\EUSO~\cite{bi:JEMEUSO}).

Intermediate steps might include:
balloon flights to test/measure some low-energy cosmic rays;
technological tests via stratospheric airplane flights; 
support activities, including: scintillation yield and Cherenkov albedo measurements;
a couple of small missions on a micro-satellite~\cite{bi:GIZMO}.

%-------------------------------------------------------------------------------
\subsection{Step 1: background measurement}
%-------------------------------------------------------------------------------

A detailed measurement/characterisation of the background, 
on the space-time scales characteristics of the EAS development is required
to improve our knowledge of it and possibly exploit it to improve background rejection.
A low-cost micro-satellite to characterise the background 
is a fundamental step to prepare such a mission~\cite{bi:ATheaPechino}.

Moreover such a micro-satellite would allow the estimation of the duty cycle
and measurements to improve the stray light control (by measuring the background
far-off nadir).

%-------------------------------------------------------------------------------
\subsection{Step 2: a technological model}
%-------------------------------------------------------------------------------

A small technological model for validation of the chosen technologies, later on,
would be required in order to: perform functional tests on critical parts of the apparatus 
(optics deployment not-to-scale, for instance);
qualify the observational approach (watching, for instance, laser shoots at the Earth);
test some technological items.

In the meantime JEM-\EUSO will provide useful information.

%-------------------------------------------------------------------------------
\section{Conclusions}
%-------------------------------------------------------------------------------

It is mandatory to clarify the Scientific objectives for post-PAO experiments.

A space-based apparatus is very challenging, also from the political and
financial point of view.

From a technical point of view a space-based apparatus is very challenging, as well.

Only a coordinated effort of the world-wide community of UHECP physicists
can hope for success.

Planning, R\&D and design should start soon, to cope with the long time-scale
required for conception, design and construction.



\end{document}